\documentclass[aps,prl,twocolumn,ap,showpacs,superscriptaddress,groupedaddress,amsmath,amssymb,epsfig]{revtex4}

\usepackage{graphicx}
\usepackage{dcolumn}
\usepackage{bm}
\usepackage{epstopdf}
\usepackage{times}
\usepackage[usenames]{color}

\newcommand{\pf}{{\phi_f}}
\newcommand{\ps}{{\phi_s}}
\newcommand{\zf}{{\zeta_f}}
\newcommand{\zs}{{\zeta_s}}
\newcommand{\cx}{{c_\mathrm{X}}}

\begin{document}

\title{Impact of single-particle compressibility on the fluid-solid phase transition for ionic microgel suspensions}

\author{M. Pelaez-Fernandez}
\affiliation{School of Physics, Georgia Institute of Technology}
\author{A. Souslov}
\affiliation{School of Physics, Georgia Institute of Technology}
\author{L. A. Lyon}
\affiliation{School of Chemistry and Biochemistry, Georgia Institute of Technology}
\author{P. M. Goldbart}
\affiliation{School of Physics, Georgia Institute of Technology}
\author{A. Fernandez-Nieves}
\affiliation{School of Physics, Georgia Institute of Technology}

\date{\today}

\begin{abstract}
We study ionic microgel suspensions composed of swollen particles for various single-particle stiffnesses.  
We measure the osmotic pressure $\pi$ of these suspensions and show that it is dominated by the contribution of free ions in solution.
As this ionic osmotic pressure depends on the volume fraction of the suspension $\phi$, we can determine $\phi$ from $\pi$, even at volume fractions so high that the microgel particles are compressed.
We find that the width of the fluid-solid phase coexistence, measured using $\phi$, is larger than its hard-sphere value for the stiffer microgels that we study and progressively decreases for softer microgels. For sufficiently soft microgels, the suspensions are fluid-like, irrespective of volume fraction. 
By calculating the dependence on $\phi$ of the mean volume of a microgel particle, we show that the behavior of the phase-coexistence width correlates with whether or not the microgel particles are compressed at the volume fractions corresponding to fluid-solid coexistence.
\end{abstract}



\maketitle

The hard sphere model (HSM), which underpins the current understanding of entropic effects in crystallization, 
applies directly to suspensions of uncharged and nondeformable spherical colloids~\cite{pusey}.
In this model, all intensive thermodynamic quantities, 
including the pressure $\pi$, depend only on the packing fraction $\phi \equiv \rho v$, 
where $\rho$ is the particle number density and $v$ is the individual particle volume.
For $\phi \ll 1$, the equation of state of the hard-sphere fluid follows the ideal gas law, $\pi \approx \rho k T = k T \phi / v$,
where $k$ is the Boltzmann constant and $T$ is the temperature, with small corrections of order $\phi^2$.
At sufficiently large $\phi$, the fluid phase (at $\pf \approx 0.49$) coexists with a crystalline solid phase (at $\ps \approx 0.54$),
so that the width of the coexistence region for this discontinuous transition $\Delta \phi_{\mathrm{HS}} \equiv \ps - \pf$ is approximately 0.05 ~\cite{alder}.

The fluid-solid phase transition is also observed in colloidal suspensions comprising microgels ~\cite{yunker,paulin,benji,chen2008,lyon}, which are deformable colloidal particles consisting of a
 network of crosslinked polymers. 
For such compressible particles, the particle volume $v$ depends on the density $\rho$, 
so that the volume fraction $\phi$ is difficult to quantify~\cite{richtering2004}.
Instead, it is convenient
to introduce the generalized packing fraction $\zeta \equiv \rho v_0$, where $v_0$ is the volume of the particle in a dilute suspension. 
For low $\rho$, $\zeta \approx \phi$; at high $\rho$, the particles are compressed, and therefore $v/v_0 = \phi/\zeta$ is significantly smaller than one. 
The overall effect of particle softness on suspension thermodynamics is determined by a combination of two features: the
elastic forces between the particles and the dependence of $v$ on $\phi$~\cite{schurtenberguer2012}.

One of the first studies of the fluid-solid phase transition of microgel suspensions reported coexistence of a fluid phase at $\zf \approx 0.59$ 
and a solid phase at $\zs \approx 0.61$~\cite{richtering}. The higher values of these quantities, compared to $\pf$ and $\ps$ for the HSM, result from the soft-repulsive interaction between the swollen microgels~\cite{kofke}. Consistent with this, the coexistence width $\Delta \zeta \equiv \zs - \zf$ was found to be smaller than $\Delta \phi_{\mathrm{HS}}$. Throughout subsequent studies, the exact values of $\zf$, $\zs$ and $\Delta \zeta$ were found to vary with the details of microgel composition~\cite{paulin, coreshell, chen2008, schurtenberguer2014, benji}. Indeed, depending on the system considered, $\Delta \zeta$ has been found to be smaller than~\cite{paulin, richtering}, similar to~\cite{schurtenberguer2014, chen2008} or larger than~\cite{coreshell, benji} 
$\Delta \phi_{\mathrm{HS}}$, without any apparent correlation with any physical characteristics of the material composing the microgel.
Evidently, the phase behavior of suspensions of soft microgels is not fully understood.

In this Letter, we address two challenges: (i) systematically quantifying the particle volume $v$, 
and therefore the packing fraction $\phi$, as a function of $\rho$ for soft microgels; and (ii) studying the effect 
of softness on the width of the coexistence region, as characterized by $\Delta \phi$. 
We accomplish this by measuring the osmotic pressure $\pi$ of ionic microgel suspensions and showing that it is dominated by the partial pressure of free ions in solution, 
which is significantly larger than the osmotic pressure due to the translational degrees of freedom of the colloidal particles.
We then calculate the packing fraction $\phi$ using a model for this ionic osmotic pressure.
We show that $\Delta \phi$ decreases for progressively softer microgels. Moreover, the softest microgels that we study
exhibit neither crystalline nor glassy states. We then connect our findings with two well-known models of soft spheres~\cite{malescio}: (i) 
the penetrable sphere model~\cite{pene}, for which $\Delta \phi > \Delta \phi_{\mathrm{HS}}$, i.e. behavior that coincides with that of the stiffer microgels in our study; and (ii)
the Hertzian~\cite{hertzian} (as well as Gaussian~\cite{gaussian}) repulsive sphere model, which exhibits a decrease in $\Delta \phi$ with increasing softness, 
such that for the softest spheres, no crystallization transition is observed.

\begin{figure}
\includegraphics{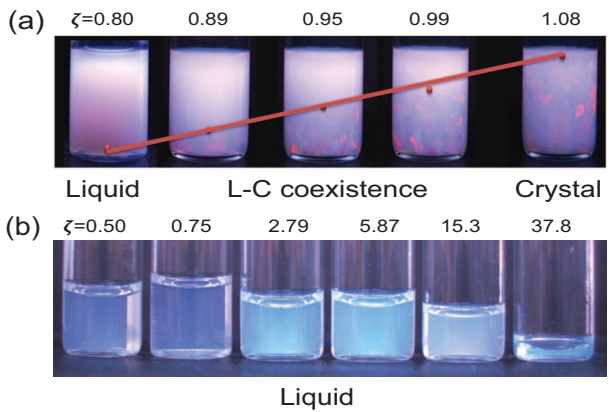}
\caption{\label{fig1} (Color online) (a) Crystal fraction versus generalized volume fraction for microgels with $\cx = 1.6$ wt\%. (b) Liquid samples at various values of $\zeta$ for microgels with $\cx = 0.2$ wt\%.}
\end{figure}

We use poly-vinylpyridine microgels crosslinked with divinylbenzene at pH 3, for which the microgels are fully swollen~\cite{vicent}. 
At this pH, the vinylpyridine groups are mostly ionized but their charge is strongly screened by the counterions in solution. 
The counterions are thus drawn inside the microgel, creating an imbalance of osmotic pressure between the inside of
the microgel and the solution outside.
Thus, electrostatic repulsions between fixed charges may be ignored but the ions exert an osmotic pressure that favors the swelling of the microgels.
This pressure is counteracted by the elasticity of the crosslinked polymer network due to the configurational entropy of the chains.
The equilibrium size of the microgel particle is determined by the balance between these two effects~\cite{macro}.

The elasticity of the polymer network is thus controlled by the crosslinker concentration $\cx$. Hence, increasing $\cx$ at a constant pH of 3 results in smaller, stiffer, swollen particles, as shown in Table I. We study the suspension phase behavior for microgels at various values of $\cx$ as a function of $\zeta$, and we visually classify whether the sample is in a fluid, solid or phase-coexisting state.
We determine $\zf$ and $\zs$, and thus $\Delta \zeta$, from a linear fit to the dependence of the crystal fraction in the samples on $\zeta$, in samples exhibiting phase coexistence~\cite{pusey, richtering}. This is shown in Fig.~1a for microgels at crosslinker concentration $\cx =1.6$ wt\%. We find that both $\Delta \zeta$ and $\zf$ increase with decreasing $\cx$, as shown in Table I, which is consistent with previous observations~\cite{benji}. Note that $\zf$ is always larger than the $\pf \approx 0.49$ seen for the HSM, suggesting that repulsive interparticle electrostatic interactions are negligible~\cite{charge}; this is a consequence of the screening of the fixed charge of the microgels by the counterions inside them. Thus, the dominant interparticle interaction arises from the elastic energy asssociated with particle deformation. Also note that suspensions composed of microgels with $\cx = 0.2$ wt\% do not exhibit the Bragg reflections indicative of a crystalline state, as shown in Fig.~1b. Hence, these suspensions remain a fluid, irrespective of $\zeta$, in agreement with previous observations on the same type of particles~\cite{benji}.

\begin{figure*}
\includegraphics{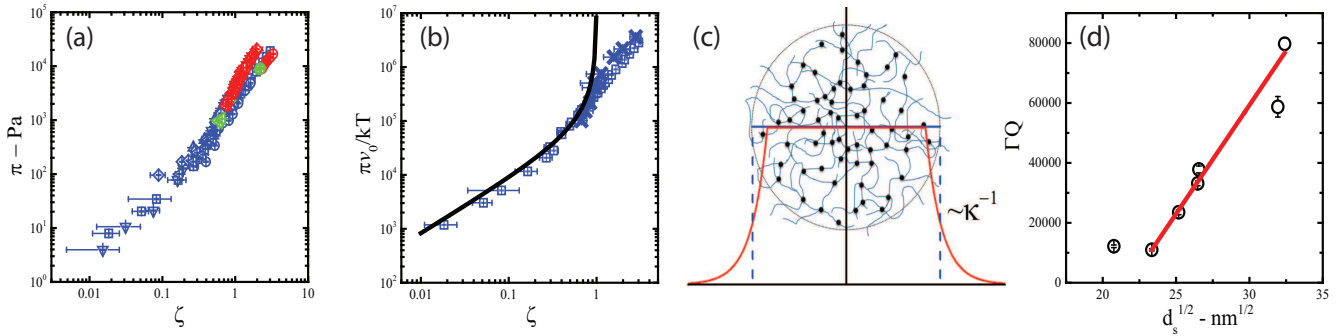}
\caption{\label{fig2} (Color online) (a) Osmotic pressure versus $\zeta$  for microgels of differing $\cx$ (see Table I for the symbol code). (b) Dimensionless osmotic pressure versus $\zeta$ for microgels with $\cx = 0.2$ wt\%. Hollow symbols: membrane osmometer; crosses: dialysis. The solid line is the best fit to the theoretical model (see text) for $\zeta < 0.63$. (c) Ionic microgel and its Donnan potential. (d) $\Gamma Q$ versus $\sqrt{d_s}$.}
\end{figure*}

To characterize the equation of state for our suspensions, we measure their osmotic pressure $\pi$ as a function of $\zeta$ using a membrane osmometer.
Significantly, $\pi(\zeta)$ does not appreciably depend on $\cx$, and therefore the particle volume $v$, as shown in Fig.~2a,
in contrast to the HSM equation of state.
Furthermore, if the pressure $\pi$ were to result from the translational degrees of freedom of the colloidal particles, at low $\zeta$ the equation of state would be described by the ideal gas law.
For $\zeta = 0.02$ this this would imply a microgel osmotic pressure of $\sim 0.1$ mPa for $\cx = 0.2$ wt\%, but
we measure $\pi \sim 5$ Pa, which is about four orders of magnitude larger.
By using dialysis~\cite{belloni}, we confirm the values of $\pi$ measured with the membrane osmometer, as shown in Fig.~2b for microgels with $\cx = 0.2$ wt\%. 
We obtained similar results for systems of microgels at other values of $\cx$.
Thus, we are confident that we have measured the osmotic pressure of the suspension correctly, 
which forces us to conclude that the dominant effect on the pressure results from some contribution not yet considered.

\begin{table}
\caption{\label{table1} Crosslinker concentration $\cx$, swollen diameter $d_s$ measured using dynamic light scattering, generalized packing fraction jump across the fluid-solid phase transition $\Delta \zeta$, and freezing generalized packing fraction $\zeta_f$.}
\begin{ruledtabular}
\begin{tabular}{l c c c}
$\cx$ (wt.$\%$) & d$_{s}$ (nm) & $\Delta \zeta$ & $\zeta_f$ \\ 
\hline
\hline
 $0.2$ {{$\Box$}} & $1050 \pm 21$ & - & - \\
 $0.5$ {{$\circ$}} & $1020 \pm 21$ & $0.41 \pm 0.07$ & $1.9 \pm 0.1 $ \\
 $1.3$ {{$\triangle$}} & $705 \pm 8$ & $0.19 \pm 0.05$ & $1.05 \pm 0.05 $ \\
 $1.6$ {{$\triangledown$}} & $701 \pm 13$ & $0.21 \pm 0.04$ & $0.87 \pm 0.04 $ \\
 $1.8$ {{$\diamond$}} & $634 \pm 8$ & $0.18 \pm 0.04$ & $0.76 \pm 0.03 $ \\
 $2.5$ {{$\lhd$}} & $545 \pm 7$ & $0.14 \pm 0.04$ & $0.65 \pm 0.04 $ \\
 $4.0$ {{$\rhd$}} & $431 \pm 6$ & $0.11 \pm 0.02$ & $0.58 \pm 0.02 $ \\

 \end{tabular}.
 \end{ruledtabular}
 \end{table}

We hypothesize that the values of $\pi$ that we measure correspond to the osmotic pressure of free counterions in solution, as there are more ions than there are microgel particles in our suspensions. To calculate this osmotic pressure $\pi_c$, we note that the Donnan potential, which confines most of the counterions to within the microgel particles, 
is constant inside a particle, but must go smoothly to zero near the particle edge over a region of thickness
corresponding to the Debye length $\kappa^{-1}$ (Fig.~2c). 
At a distance such that this potential energy is $O(k T)$, the ions are not bound to the microgel particle, and thus contribute to the osmotic pressure of the solution. 
The fraction $\Gamma$ of deconfined ions can be estimated using the model in Ref.~\cite{cloitremodel} as
 
\begin{equation}
\Gamma = \big[(\kappa^{-1} + d_s/2)^3 - (d_s/2)^3\big]/(d_s/2)^3 \approx 6 \kappa^{-1}/d_s,
\nonumber
\end{equation}

\noindent where $\kappa^{-1} \equiv \sqrt{(d_s/2)^3 / (3 \: l_B Q)}$, with $Q$ being the number of charges in a microgel, $l_B$ the Bjerrum length, and $d_s$ the diameter of a microgel: $d_s \sim \sqrt[3]{v_0}$. 
These ions occupy the region outside the microgel particles, and are sufficiently dilute that we can estimate their osmotic pressure using the ideal gas law~\cite{cloitremodel}: $\pi_c(\phi) = kT \frac{\Gamma Q}{v_{0}}\frac{\phi}{1-\phi}$. Substituting $\phi \approx \zeta$, which we find to hold for $\zeta < 0.63$, this expression fits well the $\zeta$ dependence of the osmotic pressure that we measure, and allows us to fit $\pi_c$ to experimental data and obtain the parameter $\Gamma Q$ from the fit~(see Fig.~2b). 
Furthermore, we find that the relation between the values of $\Gamma Q$ obtained from such a fit for microgels of differing $\cx$ and, therefore, differing $\sqrt{d_s}$, is consistent with the linear dependence predicted by the model, as shown in Fig.~2d. The slope of the corresponding linear fit is $(7.3 \pm 0.4) \cdot 10^3$ nm$^{-1/2}$, which compares favorably with the value estimated by using the model parameters, i.e.~$1.6 \cdot 10^3$ nm$^{-1/2}$, where we have used the value of $Q$ determined previously via titration~\cite{alberto2003}. Note that $Q$ is the same for samples of differing $c_X$ as both the deswollen size of the microgels and the particle weight fraction that results from the synthesis do not depend on cross-linker concentration \cite{vicent}.

We have thus confirmed our hypothesis that the suspension osmotic pressure is dominated by the contribution due to the counterions that lie outside of and unbound to the microgel particles. Significantly, as this contribution is sensitive to $\phi$ and not to $\zeta$, we can take advantage of the fact that $\pi =  \pi_c$ to obtain $\phi = \phi (\zeta)$. That is, we use the model to find the volume of the compressed microgels $v(\phi) = v_0 \phi/\zeta$ as a function of $\zeta$. Using this $\zeta \rightarrow \phi$ mapping, we can obtain the phase-coexistence width in terms of the microgel volume fraction $\phi$. In contrast to the behavior of $\Delta \zeta$ with $\cx$, we find that $\Delta \phi$ is approximately constant for the stiffer microgels and decreases progressively with particle softness, as shown in Fig.~3a. This decrease in $\Delta \phi$ naturally connects with the lack of crystallization for the softest microgels. We also find that $v(\phi) = v_{0}$ for volume fractions that are always above, but close to, random close packing \cite{romeo}, as shown in Figs.~3b-3h. Significantly, for the stiffer microgels the particles are not appreciably compressed within the phase coexistence region, which is indicated by square points in these figures (see Figs.~3b-3d).  In contrast, for the softer microgels, the particles are compressed in solutions that are at fluid-solid coexistence (see Figs.~3e-3g).

We now compare these experimental findings with various simulation results on models of soft spheres that interact via distinct potentials.
In such models, in addition to the entropy of the translational degrees of freedom
there is an energetic cost associated with each particle configuration, defined in terms of the pair potential $u(r)$,
such that $u(r)$ is $\epsilon (r_0 - r)^{5/2}$ (if $r < r_0$) for Hertzian, $\epsilon e^{- r^2/r_0^2}$ for Gaussian, and $\epsilon$ (if $r < r_0$) for penetrable spheres.
In all of these models, $r_0$ defines the particle radius, and $\epsilon^{-1}$ is a softness parameter. In the limit $\epsilon^{-1} \rightarrow 0$,
the Hertzian and penetrable-sphere models reduce to the HSM. [This limit is not well-defined for the Gaussian model due to the smooth tail of $u(r)$.]
For the penetrable-sphere model, the potential $u(r)$ is independent of the penetration depth $r$, such that additional overlap does not cost additional energy.
This qualitative difference distinguishes penetrable spheres from Hertzian or Gaussian spheres.

By measuring the coexistence width of our experimental system in terms of $\Delta \zeta$, we find that the width grows for softer particles, and that for the softest microgel no crystalline phase is observed.
We have not found these two features within any single numerical model of the phase behavior of soft spheres \cite{likos} -- models such as the Hertzian~\cite{hertzian} or Gaussian~\cite{gaussian} soft-sphere models exhibit no crystalline phase for sufficiently soft potentials, but also exhibit a narrowing coexistence region as the potential softens away from the hard-sphere limit; other models, such as the penetrable-sphere model~\cite{pene}, exhibit a widening of the coexistence region with increased particle softness, but in these models crystallization is observed for all values of the softness parameter.
By recomputing the width of the coexistence region in terms of $\Delta \phi$, 
we find that this width shrinks as $\cx$ decreases or, equivalently, for progressively softer particles, as shown in Fig.~3a.
Thus, for the softest microgels, models such as the Hertzian~\cite{hertzian} or Gaussian~\cite{gaussian} soft-sphere models  
seem more applicable.
However, for the stiffest microgels, we observe that $\Delta \phi > \Delta \phi_{\mathrm{HS}}$, which is not consistent with
the behavior of Hertzian or Gaussian soft spheres~\cite{hertzian,gaussian}. 
Instead, this physical aspect of stiffer microgels is better captured by the penetrable-sphere model~\cite{pene}.

\begin{figure*}
\includegraphics{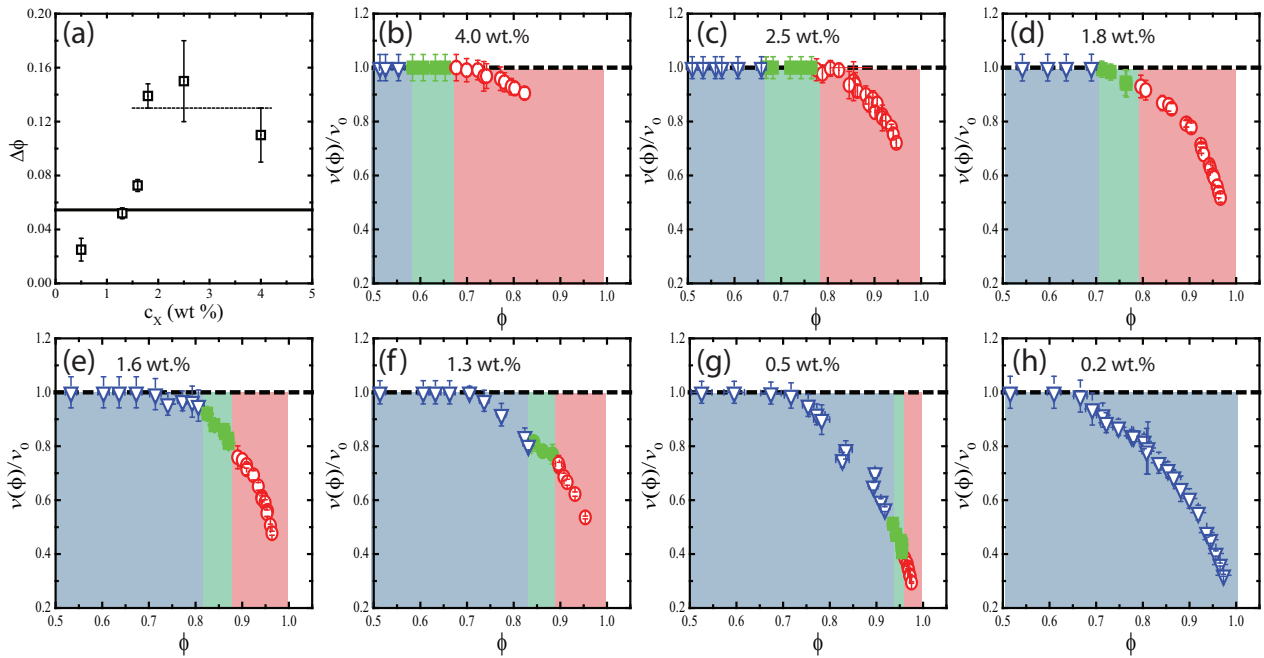}
\caption{\label{fig3} (a) Packing fraction jump at the fluid-solid phase transition for various microgel suspensions. The horizontal line shows the hard-sphere value. The dashed line reflects the approximate constancy of $\Delta \phi$ for the stiffer microgels. (b-h) Relative volume changes as a function of $\phi$ for microgels with various $\cx$ as indicated on the graphs. The symbols indicate the suspension phase: ($\triangledown$) liquid, ($\blacksquare$) coexisting phases, and ($\circ$) crystal.}
\end{figure*}

These two distinct regimes of microgel behavior may be related to the degree of particle compression at coexistence. 
In our experiments, we observe that for suspensions composed of stiffer particles (Figs.~3b-3d), there is no appreciable compression at the packing fractions corresponding to coexistence, and for these stiff particles, $\Delta \phi > \Delta \phi_{\mathrm{HS}}$ and 
$\Delta \phi$ does not depend strongly on $\cx$.
On the other hand, suspensions of softer particles (Figs.~3e-3g) exhibit appreciable compression at the packing fractions corresponding to coexistence, and for these soft particles $\Delta \phi$ is strongly dependent on $\cx$ and may be greater than, equal to, or less than $\Delta \phi_{\mathrm{HS}}$.
These two results suggest that for the stiffer microgels, where individual particles are not compressed at phase coexistence, the interaction potential is probed at distances comparable to the dilute-particle size. At these distances, the
elastic energy cost is only weakly dependent on the particle separation, as is the case in the penetrable-sphere model. This may be justified by noting that the crosslinker concentration within a microgel particle is not uniform but, rather, decreases away
from the particle center~\cite{pichot,afn}. In contrast, for the softer microgels,
where individual particles are appreciably compressed at phase coexistence, the interaction potential is probed at distances much smaller than the dilute-particle size. At these distances, the elastic energy cost is strongly dependent on the particle separation and, correspondingly, the Hertzian model better captures the thermodynamics of the coexistence region.

We can further interpret these results in terms of the bulk modulus $k_p$ of the swollen microgels, as we know that particle deswelling is only appreciable when $\pi \approx k_p$~\cite{juanjo, rubistein}. Note that the suspension osmotic pressure is comparable to $k_p$ at volume fractions that correspond to spheres at or above random close packing, indicating that below this point the assumption $\phi \approx \zeta$ is reasonable. The location in $\phi$ for phase coexistence, relative to where microgel deswelling begins, then implies that for stiffer microgels, which we relate to penetrable spheres, $\pi < k_p$ at coexistence, so that the particles are not appreciably compressed; for softer microgels, which we relate to Hertzian spheres, $\pi > k_p$ at coexistence, so that the particles are significantly compressed. In fact, for $c_X$ = 1.3\% and $c_X$ = 0.5\%, independent measurement of the single-particle bulk modulus yields \cite{sierra} $k_p = (1.6 \pm 0.1)$ kPa and $k_p = (0.40 \pm 0.02)$ kPa, respectively, with corresponding osmotic pressures at coexistence that are significantly larger; these are within the ranges $4.3 \leq \pi \leq 6.7$ kPa, for $c_X$ = 1.3\%, and $4.1 \leq \pi \leq 5.7$ kPa, for $c_X$ = 0.5\%.

Our results highlight the notion that dissolved ions play a central role in determining the osmotic pressure of colloidal suspensions, as was recently noted in sedimentation experiments on charged, nondeformable colloids~\cite{piazza, philipse}. By using a model of the ionic osmotic pressure, we obtain, for the first time, the relation between $\zeta$ and $\phi$, and this allows us to find the jump in $\Delta \phi$ between the solid and fluid phases. In this way, we find that the phase coexistence region is wider (in terms of $\phi$) than the HSM value for stiffer microgels, decreases with increasing microgel softness, and eventually disappears for sufficiently soft microgels. Our results bring coherence to a broad range of behavior previously reported for phase transformations of microgel suspensions~\cite{richtering, chen2008, coreshell} by illustrating how the particle softness determines the values of the packing fraction at which crystallization occurs and, thus, how the colloidal softness controls the width of the phase coexistence region.

We are grateful for financial support from ACS PRF 50603-DNI7, NSF DMR 12-07026, the IBB seed grant program and the Georgia Institute of Technology, and for useful discussions with  T.~C. Lubensky and A.~G. Yodh.

\end{document}